# HOSPITAL CASE COST ESTIMATES MODELLING – ALGORITHM COMPARISON

Peter Andru
Ministry of Health and Long-Term Care
Toronto, Ontario, Canada
e-mail: peter.andru@ontario.ca

Alexei Botchkarev
GS Research & Consulting
Toronto, Ontario, Canada
e-mail: alex.bot@gsrc.ca

*Abstract*

*Ontario (Canada) Health System stakeholders support the idea and necessity of the integrated source of data that would include both clinical (e.g. diagnosis, intervention, length of stay, case mix group) and financial (e.g. cost per weighted case, cost per diem) characteristics of the Ontario healthcare system activities at the patient-specific level. At present, the actual patient-level case costs in the explicit form are not available in the financial databases for all hospitals. The goal of this research effort is to develop financial models that will assign each clinical case in the patient-specific data warehouse a dollar value, representing the cost incurred by the Ontario health care facility which treated the patient. Five mathematical models have been developed and verified using real dataset. All models can be classified into two groups based on their underlying method: 1. Models based on using relative intensity weights of the cases, and 2. Models based on using cost per diem.*

## 1. INTRODUCTION

Ontario (Canada) Health System stakeholders are expecting high quality financial and clinical information for their planning and analysis purposes. All stakeholders support the idea and necessity of the consolidated, integrated and authoritative source of data and information that would include both clinical and financial characteristics of the Ontario healthcare system activities at the patient-specific level.

At the same time, generally speaking, the actual patient-level case costs in the explicit form are not available in the financial databases. Exception is for the cases submitted by the hospitals participating in the Ontario Case Costing Initiative (OCCI). However, only approximately 10% of the Ontario hospitals have been involved in the OCCI.

Financial/mathematical models should be developed in order to calculate estimates of case costs for all clinical cases information of which is stored in the Provincial Health Planning Database (PHPDB). Initial research performed by the Ontario Ministry of Health and Long-Term Care (MOHLTC) confirmed interest of the stakeholders and identified general approaches to the problem [1].

In plain language, the ultimate goal of this research effort is to assign each clinical case in the patient-specific PHPDB data warehouse a dollar value, representing the cost incurred by the Ontario health care facility which treated the patient. Five mathematical models have been developed. All models can be classified into two groups based on their underlying method: 1. Models based on using relative intensity weights of the cases, and 2. Models based on using cost per diem.

Models have been verified using real PHPDB and OCCI datasets.

Terms and definitions used in the project are presented in the Appendix 1.

## II. DESCRIPTION OF THE DATA EXPERIMENT

*A. Objective*
Design and use case cost models to calculate case cost estimates (CCE) based on the DAD database patient-specific clinical data appended with the case cost parameters published by OCCI/OCDM. Compare CCE with the case costs (benchmark) submitted to the OCCI by participating hospitals. Analyze estimating errors to verify and fine-tune the models. Logic of the experiment is shown in Fig. 1.

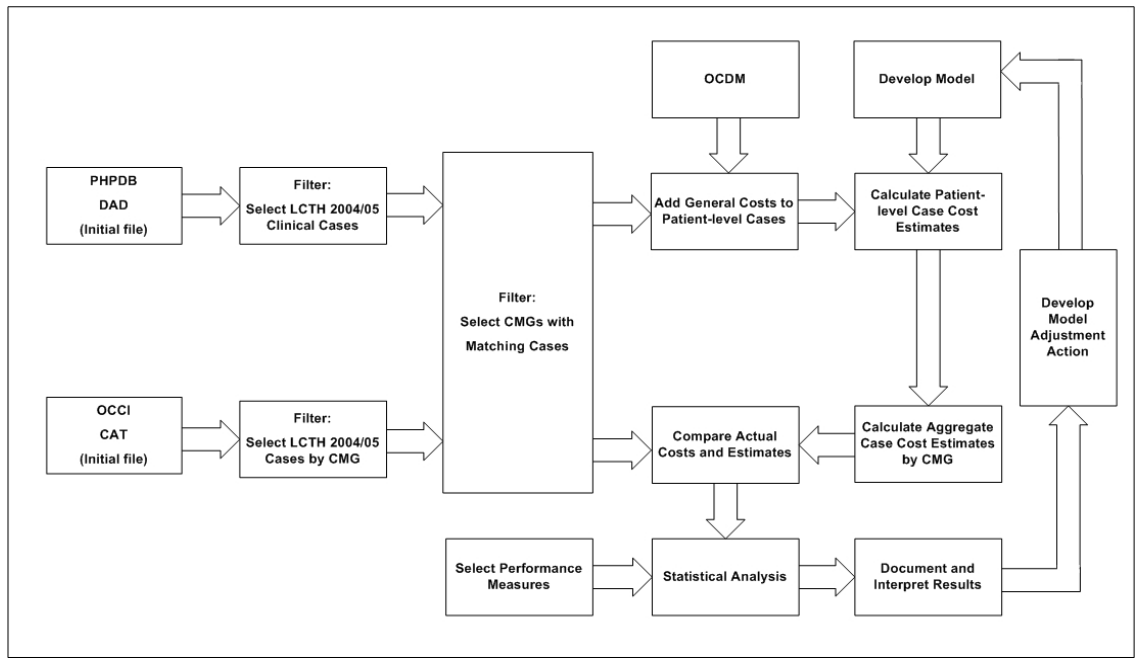

Fig. 1 Experiment Process Logic Diagram

*B. Steps of the Data Experiment Process*

1. The initial source of the patient-specific clinical data was – Inpatient Discharge (DAD) FY 2004-2005 file [2]. Dataset pertaining to a large complex teaching hospital (LCTH) 2004/05 was used in the experiment.
Layout of the file was changed to preserve only fields/columns containing clinical information which could be considered most pertinent to determining case cost estimates (e.g. CMG, diagnoses, interventions, length of stay, etc). The file has 130 fields with clinical data. It contains 7782 patient-level cases grouped in 163 CMGs (the set of data was limited to 163 CMG codes with matching number of cases between DAD and OCCI CAT (Cost Analysis Tool – www.occp.com – Costing Reports).
2. Clinical information in the file was appended with the following OCDM cost data for LCTH hospital [3]: Acute expenses;
Acute Weighted Cases; Cost per Weighted Case; Acute Direct Cost per Diem; Acute Overhead Cost per Diem; Acute Total Cost Per Diem; Acute Total Patient Days.
3. Several (total of 5) financial models were developed and used one at a time to calculate CCE for each patient-level case.
4. Using CCEs and simulating the query: "What's the total cost of the LCTH CMG xyz cases in 2004-05?", aggregate cost estimates (and several other parameters) were calculated for each of the 163 CMG codes:
- Estimate of the Total cost of the cases in the CMG group
- Estimate of the Average cost of the case in the CMG group
- Estimate of the Standard deviation of the case costs in the CMG group
- Estimate of the Minimum case cost in the CMG group
- Estimate of the Maximum case cost in the CMG group
The reason for selecting and calculating estimates of the above parameters is that the same parameters are available from the OCCI CAT database [4].
5. OCCI CAT database was queried: "What are the case cost parameters of the LCTH CMG xyz cases in 2004-05?" for each of the 163 CMG codes. OCCI CAT returns the same case cost parameters as outlined in Step 4. The difference is that this is the official OCCI data, and it was used later as the Actual (correct, verified) cost.
6. Estimated (step 4) and actual (step 5) cost parameters were compared and estimating errors characterizing the efficiency of the financial models were calculated for each of the 163 CMG codes:
- Error estimating the Total cost of the cases in the CMG group

$$E(ACCE_{cmg}) = ACCE_{cmg} - AACC_{cmg}$$

- Error estimating the Average cost of the case in the CMG group

$$E(avgCCEcm) = avgCCEcmg - avgACCcmg$$

- Error estimating the Standard deviation of the case costs in the CMG group

$$E(stdevCCEcmg) = stdevCCEcmg - stdevACCcmg$$

- Error estimating the Minimum case cost in the CMG group

$$E(minCCEcmg) = minCCEcmg - minACCcmg$$

- Error estimating the Maximum case cost in the CMG group

$$E(maxCCEcmg) = maxCCEcmg - maxACCcmg$$

7. Estimating error analysis was conducted to compare various financial models, determine their accuracy, identify factors impacting results, and fine-tune models. Analysis of all types of errors calculated on step 6 would require handling matrixes with dimensions 5 x 163. Being feasible generally, it wouldn't provide results which could be easily understood and interpreted. That led to the formulation of a limited number of performance measures.

The first performance measure is Pab - the percentage of the CMG groups for which the absolute value of the relative error estimating total cost of cases per CMG group lies within certain limits.

$$P_{ab} = n_{ab} / n,$$

where

n – number of the CMG groups in the experiment (in this case – 163);

$n_{ab}$ – number of the CMG groups for which the ratio of the estimating error of the total cost of cases for the CMG group to the actual total cost for this CMG group (in percent) falls within the interval from a to b:

$$n_{ab} \subset n \left\{ a < \left| \frac{E(ACCEcmg)}{AACCcmg} \right| \leq b \right\}$$

Three other measures of performance characterize absolute error of estimating average, minimum and maximum case costs per CMG code averaged over all (163) CMG codes:

$$\bar{E}(avgCCEcmg) = \left\{ \sum_n |E(avgCCEcmg)| \right\} / n \quad \bar{E}(minCCEcmg) = \left\{ \sum_n |E(minCCEcmg)| \right\} / n$$

$$\bar{E}(maxCCEcmg) = \left\{ \sum_n |E(maxCCEcmg)| \right\} / n$$

**III. FINANCIAL MODELS and DATA ANALYSIS**

*A.    PAC Relative Intensity Model – Model 1*

$$CCE_i = PAC\_RIW_i * CPWC,$$

where:

$PAC\_RIW_i$ – relative intensity of the i-th case. Parameter is available in the DAD database – field $PAC\_RIW\_WT$.

$CPWC$ – cost per weighted case. Parameter is calculated annually by OCDM for each hospital. For the LCTH in 2004-05 $CPWC \sim \$6,000$.

Performance measures for the model are given in Table 1, which combines measures for all models (see Section IV. Data Experiment Results and Conclusions).

Large segment of the estimates of the total cost of cases per CMG ($ACCE_{cmg}$) is relatively accurate – for 68% of the CMG groups relative error is under 20%. At the same time, for 18% of the CMG groups relative error is over 30%.

It was noticed that for a number of CMG codes, PAC RIW weights were set at exactly the same value for all cases in the group. Although this situation is generally possible, in some instances it raises questions. For example, for the CMG group 232 which contains 9 patient-level cases all PAC RIWs are set at 0.9002. Length of stay for the cases in this group varies from 1 to 20 days, and OCCI database indicates that minimum and maximum case costs in the group are $821.81 and $17,651.78, respectively. There're 12 CMG groups in the dataset with this type of data.

Related observation is that the accuracy of the cost estimates for the CMG groups in question are usually very low.

Estimates of the average case costs are more accurate than estimates of the minimum and maximum case costs.

More detailed error analysis shows that within the CMG group cases with lowest costs are mostly overestimated. At the same time, cases with the highest costs are mostly underestimated (in many cases – significantly). The figure below illustrates the biased nature of minimum and maximum cost estimates calculated with this model.

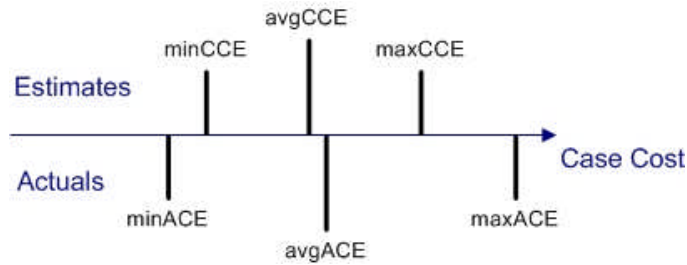

Fig. 2 Biased minimum and maximum case cost estimates

Minimum and maximum costs are usually attributed to the atypical cases (with either very short or very long length of stay).

The overall comment is that the PAC Relative Intensity Model, although being simple, straightforward and providing relatively accurate estimates for the typical cases, calculates biased results for atypical cases and requires verification of the PAC RIW weights for the single value CMG groups.

### B. PAC Relative Intensity Model (RIW-modified) – Model 2

As it was noted in the description of the Model 1, it was observed that in some instances PAC weights are set identical for all cases across CMG group. That results in large errors. Current model was intended to alleviate this drawback by replacing the identical PAC weights (PAC_RIW_WT) with normalized RIWval for the affected CMG codes. Normalized weights were used in order to keep the total weight of the CMG group at the same PAC weight value.

PAC Relative Intensity Model (RIW-modified) can be formulated the following way:

$$CCE_i = PAC\_MOD_i * CPWC,$$

where:

$$PAC\_MOD_i = \begin{cases} PAC\_RIW_i, \text{ if } minPAC\_RIW_i \neq maxPAC\_RIW_i \text{ for n-th CMG group} \\ \qquad \text{or } minPAC\_RIW_i = maxPAC\_RIW_i \text{ and } minRIW\_val_i = maxRIW\_val_i \text{ for n-th CMG group} \\ RIW\_val_i *(\sum_{cmg_n} PAC\_RIW_i / \sum_{cmg_n} RIW\_val_i), \text{ if } minPAC\_RIW_i = maxPAC\_RIW_i \text{ for n-th CMG group} \end{cases}$$

$PAC\_RIW_i$ – relative intensity of the i-th case. Parameter is available in the DAD database – field PAC_RIW_WT;

$RIW\_val_i$ – relative intensity of the i-th case calculated by the CIHI. Parameter is available in the DAD database – field $RIW_{val}$;

$minRIW\_val_i$ and $maxRIW\_val_i$ – minimum and maximum values of the $RIW\_val_i$ for the n-th CMG group;

$minPAC\_RIW_i$ and $maxPAC\_RIW_i$ – minimum and maximum values of the $PAC\_RIW_i$ for the n-th CMG group;

CPWC – cost per weighted case. Parameter is calculated annually by the OCDM for each hospital. For the LCTH in 2004-05 CPWC~$6,000.

RIW-modified model impacted PAC weights of the following CMG groups: 232, 393, 767, and 884. The table below shows that the accuracy of estimating minimum and maximum costs for the CMG groups involved has been improved.

TABLE II
RIW-modified Model Performance Measures for Select CMGs

| CMG | | $E(\text{stdevCCE}_{cmg})$ | $E(\text{minCCE}_{cmg})$ | $E(\text{maxCCE}_{cmg})$ |
|---|---|---|---|---|
| 232 | Non-modified | -5,273.05 | 4,573.99 | -12,255.98 |
| | Modified | -2,724.45 | 1,216.33 | -9,586.52 |
| 884 | Non-modified | -6,370.07 | 8,331.54 | -7,797.72 |
| | Modified | -2,504.95 | 3,964.70 | -2,196.52 |

Due to the fact that the total sum of the PAC weights per CMG was not changed (because of normalization), estimates of the total cost of the cases in the CMG group and average cost of the case in the CMG group and their respective errors were not affected.

Performance measures for the model are given in Table II, which combines measures for all models (see Section IV. Data Experiment Results and Conclusions).

Some improvement may be noticed as compared to the Model 1, but it's relatively small because only four out of 163 CMG groups in the dataset were affected.

For the following CMG groups – 434, 480, 551, 771, 778, 790, 883 – both PAC and RIW weights are set to a single value. These cases could not be improved with Model 2. But they should be given attention later as the estimating errors for them are significant. E.g. for the CMG 551 all case costs estimates are at the same level - $4,5K (as the PAC and RIW weights are the same), but OCCI minimum and maximum costs in the group are 1K and 10K respectively.

C.  *PAC Relative Intensity Model (RIW-modified, total expenses normalized) – Model 3*

According to the project's objectives precautions are to be taken to assure that the data generated as a result of the project does not contradict to the data currently published by OCCI/OCDM. One of the important figures to preserve intact is the total acute expenses of the hospital. Due to the limited scope of the data used in the experiment it can not be done directly, but measures are to be taken to make sure that the estimated total cost of the cases in the dataset equals to the total cost of these cases in the OCCI database, i.e. the following equation should hold:

$$\sum_{cmg_n} \sum_{i=1,\ldots,N_{c,cmg}} \text{CCE}_i = \sum_{cmg_n} \text{avgACC}_{cmg} * N_{c,cmg}$$

In the real life situation, the right-hand side of the expression equals to the total acute expenses of the hospital annually published by the OCDM.

PAC Relative Intensity Model (RIW-modified and total expenses normalized) can be formulated the following way:

$$\text{CCE}_i = \text{PAC\_MOD}_i * \text{CPWC} * \left\{ \sum_{cmg_n} (\text{avgACC}_{cmg} * N_{c,cmg}) / \sum_{cmg_n} \sum_{i=1,\ldots,N_{c,cmg}} \text{PAC\_MOD}_i * \text{CPWC} \right\}$$

where:

$$\text{PAC\_MOD}_i = \begin{cases} \text{PAC\_RIW}_i, \text{ if } \min\text{PAC\_RIW}_i \neq \max\text{PAC\_RIW}_i \text{ for n-th CMG group} \\ \quad \text{or } \min\text{PAC\_RIW}_i = \max\text{PAC\_RIW}_i \text{ and } \min\text{RIW\_val}_i = \max\text{RIW\_val}_i \text{ for n-th CMG group} \\ \text{RIW\_val}_i * \left( \sum_{cmg_n} \text{PAC\_RIW}_i \Big/ \sum_{cmg_n} \text{RIW\_val}_i \right), \text{ if } \min\text{PAC\_RIW}_i = \max\text{PAC\_RIW}_i \text{ for n-th CMG group} \end{cases}$$

$\text{PAC\_RIW}_i$ – relative intensity of the i-th case. Parameter is available in the DAD database – field PAC_RIW_WT;

$\text{RIW\_val}_i$ – relative intensity of the i-th case calculated by the CIHI. Parameter is available in the DAD database – field $\text{RIW}_{val}$;

$\min\text{RIW\_val}_i$ and $\max\text{RIW\_val}_i$ – minimum and maximum values of the $\text{RIW\_val}_i$ for the n-th CMG group;

$\min\text{PAC\_RIW}_i$ and $\max\text{PAC\_RIW}_i$ – minimum and maximum values of the PAC_RIWi for the n-th CMG group;

$\text{CPWC}$ – cost per weighted case. Parameter is calculated annually by the OCDM for each hospital. For the LCTH in 2004-05 $\text{CPWC} \sim \$6,000$;

$\text{avgACC}_{cmg}$ - average actual cost of the case in the CMG group. Parameter is available in the OCCI database.

Performance measures for the model are given in Table II, which combines measures for all models (see Section IV. Data Experiment Results and Conclusions).

Normalization didn't lead to any significant changes in accuracy, but technically it's a required step to assure compliance with the OCDM/OCCI data.

This step will be included in all other models as a default.

### D. Cost per Diem Model – Model 4

Model 4 multiplies length of stay (in days) by the cost per diem. The third multiplier is used for normalization of the total cost.

$$\text{CCE}_i = \text{LOS}_i * \text{CpD} * \left\{ \sum_{cmg_n} (\text{avgACC}_{cmg} * N_{c,cmg}) \Big/ \sum_{cmg_n} \sum_{i=1,\ldots,N_{c,cmg}} \text{LOS}_i * \text{CpD} \right\}$$

where:

$\text{LOS}_i$ – Length of Stay for the i-th case. This parameter is available in the DAD database – field TotalL.

$\text{CpD}$ – cost per diem. Parameter is calculated annually by OCDM for each hospital. For the LCTH in 2004-05 $\text{CpD} \sim \$1,600$. Note: Total cost per diem is used without breakdown into Direct and Indirect costs which are also available in the OCDM report.

Performance measures for the model are given in Table II, which combines measures for all models (see Section IV. Data Experiment Results and Conclusions).

The model provides very poor estimates for the total cost per CMG group – for 57% of the groups the estimating error exceeds 30%.

Estimates of the maximum costs are at about the same level of accuracy as for other models. The error is large and costs are mostly underestimated (although the bias is not as consistent as in the relative intensity models).

Important observation is that estimates of the minimum costs have much better accuracy than of the other models.

### E. Cost per Diem Model with the Type of Stay Differentiation – Model 5

Model 5 elaborates on Model 4 by explicitly calculating costs of the three different types of the patient's stay in the hospital: in acute care, in alternate care and in a special care unit. Costs for each type of stay are set in relation to the average cost per diem $CpD$.

Cost per Diem Model with the Type of Stay Differentiation can be formulated the following way (the third multiplier – in brackets - is used for normalization of the total cost):

$$CCE_i = \left\{ (LOSac_i - LOSsc_i/24) * K1 + LOSalc_i * K2 + (LOSsc_i/24) * K3 \right\} * CpD *$$

$$\left\{ \sum_{cmg_n} (avgACC_{cmg} * N_{c,cmg}) \bigg/ \sum_{cmg_n} \sum_{i=1,\ldots,N_{c,cmg}} \left\{ (LOSac_i - LOSsc_i/24) * K1 + LOSalc_i * K2 + (LOSsc_i/24) * K3 \right\} * CpD \right\}$$

where:

$LOSac_i$ – Length of Stay in acute care for the i-th case. This parameter is available in the DAD database – field $AcuteL$.

$LOSalc_i$ – Length of Stay in alternate care for the i-th case. This parameter is available in the DAD database – field $ALClen$.

$LOSsc_i$ – Length of Stay in a special care unit for the i-th case. This parameter is available in the DAD database – field $TotHRS$.

$CpD$ – cost per diem. Parameter is calculated annually by OCDM for each hospital. For the LCTH in 2004-05 $CpD \sim \$1,600$. Note: Total cost per diem is used without breakdown into Direct and Indirect costs which are also available in the OCDM report.

$CpD * K1$ – cost per diem in acute care. $K1 > 1$.

$CpD * K2$ – cost per diem in alternate care. $K2 < K1$.

$CpD * K3$ – cost per diem in a special care unit. $K3 > K1$.

Note: If $K1=K2=K3$ Model 5 transforms into Model 4. Actual value is irrelevant due to the normalization.

Values for the coefficients $K1, K2, K3$ were selected by running the model and minimizing errors. The following values were selected: $K1=1.3, K2=0.5, K3=2.85$. Although these values minimize errors and look reasonable, they are to be verified on the larger datasets.

Performance measures for the model are given in Table II, which combines measures for all models (see Section IV. Data Experiment Results and Conclusions).

Performance measures show that Model 5 provided some enhancement in the average and minimum case cost estimates and more significantly to the maximum case cost estimates. Still, due to the high level of errors the total case cost per CMG is estimated inappropriately.

### IV. DATA EXPERIMENT RESULTS AND CONCLUSIONS

Performance measures for the models with optimal coefficients are presented in the table below.

TABLE II
Financial Models Performance Measures

| Model Number | | 1 | 2 | 3 | 4 | 5 |
|---|---|---|---|---|---|---|
| $\bar{E}(\text{avgCCE}_{\text{cmg}})$, $ | | 2.0 | 2.0 | 2.0 | 4.0 | 3.3 |
| $\bar{E}(\text{minCCE}_{\text{cmg}})$, $ | | 4.1 | 4.0 | 4.1 | 1.1 | 1.0 |
| $\bar{E}(\text{maxCCE}_{\text{cmg}})$, $ | | 27.0 | 26.9 | 26.9 | 28.4 | 20.9 |
| $P_{ab}$, % | | | | | | |
| a | b | | | | | |
| 0 | 5 | 20 | 20 | 18 | 7 | 8 |
| 5 | 10 | 22 | 22 | 23 | 6 | 7 |
| 10 | 15 | 13 | 13 | 13 | 6 | 10 |
| 15 | 20 | 13 | 13 | 14 | 6 | 6 |
| 20 | 30 | 15 | 15 | 14 | 18 | 18 |
| 30 | 50 | 11 | 11 | 11 | 26 | 26 |
| 50 | | 7 | 7 | 7 | 31 | 25 |

Results of the data experiment show feasibility of the selected approach.

All models can be classified into two groups based on their underlying method:

1. Models based on using relative intensity weights of the cases. This group includes Models 1, 2, 3.

2. Models based on using cost per diem. This group includes Models 4, 5.

Despite the differences of the individual models, each group has certain common features. The first group models display relatively good performance estimating average CMG costs, and hence the total case cost of the CMG group, but they are much less accurate estimating minimum and maximum CMG case costs. On the other hand, the second group shows good estimating performance of the minimum and maximum case costs, but fails in estimating the total costs of the CMGs.

The above observation prompts an approach for the next phase of the research. Future algorithms will combine estimates calculated by the models belonging to the both groups. On the first step, two models (e.g. Model 3 and Model 5) will be used to calculate all estimates. On the second step, minimum and maximum case cost estimates of the each CMG group of the relative intensity model (Model 3) will be replaced with minimum and maximum case cost estimates calculated with the cost per diem model (Model 5).

Also, the values of the empirical coefficients K1 - K3 were selected by a manual trial and error method. That limited the range of possible combinations of coefficients and increased probability of missing the optimum. To overcome this deficiency, an Excel macro will be developed, which would allow to automatically "scan" thousands of combinations of coefficients within reasonable time.

The following set of criteria will be used to optimise model's performance:

- Minimum of large errors, i.e. percentage of the CMG groups with errors in the intervals from 30 to 50% and more than 50%.

- Minimum of very large errors, i.e. percentage of the CMG groups with errors more than 50%.

- Maximum of the small errors, percentage of the CMG groups with errors in the intervals from 0 to 5% and from 5 to 10%.

Sequence of applying criteria is important.

## ACKNOWLEDGMENT

The work was prepared in connection with both authors' official duties at the Ontario Ministry of Health and Long-Term Care.

**Appendix 1. Terms and definitions**

**Case**. An instant of a disease that led to the individual's inpatient stay, which has been registered by the health service organization, reported to the Canadian Institute for Health Information (CIHI) DAD, database and eventually has a corresponding record with all appropriate attributes in the PHPDB.

**Case Cost.** Expenditures (direct and indirect) incurred by the health service facility relating to the treatment of a specific (patient-level) case. There's no way of "precise measuring" of the dollar value of each and all specific cases. Irrespective of the methodology and calculation techniques employed to determine the case cost, dollar value of a specific case is always an approximation of the "real-life" hospital expenses. In some sense, case cost is a conceptual phenomenon reflecting hospital business functioning as a clinical and financial entity.

**Case Costing.** A process of allocating expenditures of various hospital departments to each individual case with an objective to determine the case cost. The process is performed according to the Ontario Cost Distribution Methodology (OCDM) and OCCI methodology. This project is not involved in the case costing process, its formal description, implementation or modification. Although, the project uses case costing results (case costs), and parameters employed by the case costing methodology. In general, case costing can be accomplished in various ways. Within the framework of this project, the term case costing is used only with the implication that the process is based on the OCDM/OCCI methodology.

**Actual Case Cost (ACC).** Case Cost that is considered to be the most trustworthy and most closely reflecting "real-life" hospital expenditures on a specific case is referred to as Actual Case Cost. Actual Case Costs are used as a benchmark for comparing various case costs.

**Case Cost Model (CCM).** A formal representation of the Case Cost, as a conceptual phenomenon, which involves mathematics, logical expressions, well-defined procedures, computer software, that is constructed with the purpose of producing output – Case Cost Estimate, as a function of one or more clinical and financial parameters. CCM can constitute a single formulae/equation or a set of complex algorithms implemented in a software package.

**Case Cost Estimate (CCE).** Approximate dollar value of the Case Cost, determined/calculated by employing the Cost Case Model. CCE may constitute a single value or a range of values.

**Aggregate Actual Case Cost (AACC).** The sum of two or more actual costs of cases (ACCs) usually produced as a result of a query with the underlying question such as: What was the actual cost of all cases of the XYZ hospital? What was the actual cost of all cases with Case Mix Group (CMG) code xxx?

**Aggregate Case Cost Estimate (ACCE).** The sum of two or more cost estimates of cases (CCEs) usually produced as a result of a query with the underlying question such as: What was the cost estimate of all cases of the XYZ hospital? What was the cost estimate of all cases with CMG code xxx?

**Cost Modelling.** 1. A process of applying structured methodology to create and validate Case Cost Models (CCM). 2. A process of employing CCM to produce Case Cost Estimate (CCE).

**Financial Modelling.** A complex process which involves people, data, equipment, software, methods, and includes a set of interrelated activities of: acquiring clinical and financial data; creating and validating Case Cost Models and employing them to produce Case Cost Estimates; extracting, transforming and loading data; performed to integrate clinical and financial data in the PHPDB and make it available to the healthcare analysts.